\documentclass[12pt]{article}
\usepackage{graphicx}
\usepackage{amsmath}
\usepackage{feynmp}

\newcommand{\Z}{{\cal Z}}
\renewcommand{\S}{{\cal S}}
\begin{document}

\begin{flushright}
HET--1120\\
TA--558
\end{flushright}

\begin{center}
{\large Numerical Field Theory on the Continuum}\\[12pt]
Stephen C. Hahn and G. S. Guralnik\\
{\it Department of Physics, Brown University,}\\
{\it Providence RI 02912--1843}
\end{center}

\begin{abstract}
An approach to calculating approximate solutions to the continuum
Schwinger--Dyson equations is outlined, with examples for $\phi^4$ in $D=1$.
This approach is based on the source Galerkin methods developed by Garc\'\i a,
Guralnik and Lawson.  Numerical issues and opportunities for future calculations
are also discussed briefly.
\end{abstract}

\baselineskip=21pt
\section{Introduction}

The now-conventional technique of numerical calculation of
quantum field theory involves evaluating the path integral
\begin{equation}
\Z(J) = \int_{\Gamma} d\phi\,  \exp\left[-\S(\phi) + J \phi\right]
\label{complexcont}
\end{equation}
on a spacetime lattice using Monte Carlo integration methods.  Monte Carlo
methods have been successful for an interesting class of problems; however,
these techniques have had less success in evaluating theories with actions
that are not manifestly positive definite or which have important effects from
the details of fermionic interactions beyond the quenched approximation.  In
these cases, we have either a basic algorithmic difficulty or an  inadequacy
of compute power.

We are developing an alternative computational method which works both on the
lattice and the continuum and which handles fermions as easily as bosons.
Furthermore, our ``source Galerkin'' method is less restrictive as to the class
of allowed actions.  Source Galerkin tends to use significantly less compute
time than Monte Carlo methods but can consume significantly larger amounts of
memory.%
\footnote{This can happen because the iterative process for improving a source Galerkin
calculation involves successively higher-point Green functions,
whereas calculation of these correlations in a Monte Carlo scenario is
optional.}
This talk is confined to continuum applications;  examples of lattice
calculations have been given elsewhere
\cite{group:initial,group:lawson:bosons,group:lawson:fermions}.  While it is not
clear that Source Galerkin can replace Monte Carlo techniques, it appears that
it will be able to solve some problems which are currently inaccessible.

Our approach begins with the differential equations satisfied by the
vacuum functional $\Z$ for a quantum field theory with external sources.
For the sake of simplicity, in this talk we will mostly confine our
attention to $\phi^4$ interactions. To date, we have also studied non-linear
sigma models and four-fermion interactions and have gauge theory
calculations in progress.

The vacuum persistence function $\Z$ for a scalar field $\phi$ with
interaction $g\phi^4/4$ coupled to a scalar source $J(x)$ satisfies
the equation:
\begin{equation}
\left( -\partial^2_D + M^2 \right) {\delta \Z \over \delta J(x)}+
g\,{{\delta}^3 \Z \over \delta {J}^{3}(x)} \,-\,J(x) \Z \,=\,0
\label{euc1}
\end{equation}
The source Galerkin technique is designed to directly solve functional
differential equations of this type. Before we proceed to outline a
solution technique, it is essential to point out that this equation by
itself does not uniquely specify a theory \cite{group:ultralocal}.
This is dramatically illustrated by considering the special case of
the above equation limited to one degree of freedom (zero dimensions).
\begin{equation} 
M^2 {d\Z \over dJ} + g {d^3\Z \over dJ^3} - J\Z = 0
\end{equation} 
This is a third order differential equation, and therefore possesses three
independent solutions.  It is easy to see that one of the solutions for small
$g$ asymptotically approaches the perturbation theory solution, while the second
asymptotically approaches the loop expansion ``symmetry breaking'' solution for
small $g$ and the third solution has an essential singularity as $g$ becomes
small. The situation becomes much more interesting for finite dimensions where
the infinite class of solutions coalesces or becomes irrelevant in a way which
builds the phase structure of the field theory. Any numerical study must be
cognizant of the particular boundary conditions and hence solution of the class
of solutions that is being studied.  Care must be taken to stay on the same
solution as any iterative technique is applied.  The solutions discussed in this
talk will be the usual ones which correspond to the symmetry preserving
solutions obtained from evaluating a path integral with real axis definitions
for the regions of integration. These are the solutions that are regular in the
coupling, $g$, as it approaches zero.

We, for convenience, write (\ref{euc1}) in the form:
\begin{equation}
\hat E_j \Z(j) = 0
\end{equation}
The source Galerkin method is defined by picking an approximation
$\Z^*(j)$ to the solution $\Z(j)$ such that
\begin{equation}
\hat E_j \Z^*(j) = R
\end{equation}
where $R$ is a residual dependent on $j$. We pick the parameters of our approximation
to make
this residual as small as possible on the average. To give this
statement a meaning, we define an inner product over the domain of
$j$: {\it i.e.\/} $(A,B)\equiv\int d\mu(j)\,A(j)B(j)$.  In addition we
assume we have a collection of test functions which are members of a
complete set) $\{\varphi_i(j)\}$ The source Galerkin minimization of
the residual $R$ is implemented by setting the parameters of our test
function $\Z^*(j)$ so that projections of test functions against the
residual vanish so that $|| \Z^* - \Z ||_2 \to 0$ as the number of
test functions $\to\infty$.

The equations defining the quantum field theory are differential
equations in the field sources and spacetime. While it is
straightforward to deal with the spacetime problem by resorting to a
lattice, we can remain in a continuum formulation by taking advantage if our
knowledge of functional integration.  We know how to evaluate Gaussian
functional integrals on the continuum:
\begin{multline} 
\int [dj] \exp\left[ \int_{xy}j(x)A(x,y)j(y) +
\int_{x}j(x)\beta(x) \right] \\ 
= \frac{1}{\sqrt{\det A}}\exp\left[
\int_{xy}\beta^*(x)A^{-1}(x,y)\beta(y) \right] 
\end{multline}
Consequently, we can evaluate integrals of the form:
\begin{equation}
I = \int [dj] \exp\left[ -j^2(x)/\epsilon^2 \right] P(j)
\end{equation}

Using this we can define an inner product of sources on the continuum
as follows:
\begin{equation}
(j(x_1)\cdots j(x_n), j(y_1)\cdots j(y_m))_j = \begin{cases}
\epsilon^{n+m}\delta_+\{x_1\cdots x_n y_1 \cdots y_m\}&n+m\ {\rm even}\\
0&{\rm otherwise}
\end{cases}
\end{equation}
where we have absorbed a factor of 2 by redefining $\epsilon$.
$\delta_+$ is defined by
\begin{gather}
\delta_+\{x\alpha\beta\cdots\} =
\delta(x-\alpha)\delta_+\{\beta\cdots\} +
\delta(x-\beta)\delta_+\{\alpha\cdots\} + \cdots,\\
\delta_+\{x\alpha\} = \delta( x-\alpha ).
\end{gather}

In addition to this inner product definition, we need good guesses for
approximate form for $\Z^*$ and numerical tools to calculate,
symbolically or numerically, various functions and their integrals,
derivatives, and so on.  For most of our calculations we have found it
very useful to choose a lesser known class of functions, with
very suitable properties for numerical calculation, known as Sinc functions.
We take our notation for the Sinc functions from Stenger
\cite{mono:stenger:sinc-methods}:
\begin{equation}
S(k,h)(x) = \frac{\sin( \pi(x-kh)/h)}{\pi(x-kh)/h}
\end{equation}
Some of the identities that Sinc approximations satisfy are given below: 
\begin{gather}
S(k,h)(lh) = \delta_{kl}\\
\int_x S(k,h)(x) S(l,h)(x) = \delta_{kl}\\
\int_x F(x) \approx h \sum_{k=-N}^N F(kh)\\
F(x) \approx \sum_{k=-N}^N F(kh)S(k,h)(x)\\
F'(x) \approx \sum_{k=-N}^N F(kh)S'(k,h)(x)\\
F^{(n)}(lh) \approx \sum_{k=-N}^N F(kh)\delta^{(n)}_{k-l}
\end{gather}
These properties, which are proven and expanded upon greatly in
\cite{mono:stenger:sinc-methods}, make these functions very easy to use for
Galerkin methods, collocation, integration by parts, and integral equations.

\begin{figure}
\begin{center}
\includegraphics[height=2in]{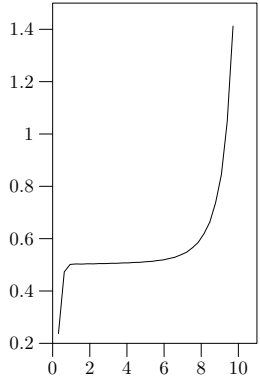} \caption{Mass,
$m^*$, versus distance, $x$, for two-dimensional free scalar field.
Note breakdown near origin (approximation of $\delta$-function) and at
large distance (spatial truncation). $m_0 = 0.5$.}
\label{fig:free-two}
\end{center}
\end{figure}

With the definition of a norm and set of expansion functions, we can
postulate an ans\"atz for ${\cal \Z}$
\begin{equation}
{\cal \Z}^* = \exp\left[\sum \int_{xy} j(x)G_2(x-y)j(y) + \cdots\right]
\end{equation}
where each Green function, $G_n$, is represented by a $d$-dimensional
Sinc expansion.  For $d=4$:
\begin{multline}
G_2(x-y) = \sum
G_2^{ijkl}\\
S(i,h)(x^0-y^0)S(j,h)(x^1-y^1)S(k,h)(x^2-y^2)S(l,h)(x^3-y^3)
\end{multline}
It is easy to examine this expansion in the case of a free field where
we limit the approximation to the terms quadratic in $j(x)$. While the
example is trivial, it shows that, as always, numerical approximations
must be handled with care.  Results of this calculation are shown in Figure
\ref{fig:free-two}.

This straightforward expansion works for interacting theories but
with more than $G_2$, the computational costs and storage costs become
overwhelming: $G_{2n}$ requires $N^{(2n-1)d}$ storage
units. There are many ways that storage costs can be reduced, but in
general these approaches are difficult, not particularly elegant, and eventually
reaches a limit due to the exponential growth in the number of
coefficients of the representation.

We can use our knowledge of the spectral representations of field
theory and graphical approaches to introduce a much more beautiful and
intuitive approach to producing candidates for $\Z^*$.  We introduce
regulated Lehmann representations. These build in the appropriate
spacetime Lorentz structure into our approximations and remove the
growth of operational cost with spacetime dimension shown by our
previous na\"\i ve decomposition into complete sets of functions.  Any
exact two-point function can be represented as a sum over free two-point
functions. We choose as the basis of our numerical solutions, a regulated
Euclidean propagator structure: 
\begin{align} \Delta(m;x) \equiv
\int (dp) \frac{e^{ip\cdot x - p^2/\Lambda^2}}{p^2+m^2} &= \int (dp)
\int_0^\infty ds\, e^{ip\cdot x - p^2/\Lambda^2 - s(p^2+m^2)}\\ &=
\frac{1}{(2\pi)^d}\int_0^\infty
ds\,\left[\frac{\pi}{s+1/\Lambda^2}\right]^{d/2}e^{-sm^2 -
\frac{x^2}{4(s+1/\Lambda^2)}} 
\end{align}
This regulation assures that we never have to deal with infinities in any
calculated amplitude as long as we keep the cutoff finite.  

This integral can be approximated using Sinc methods
\begin{equation}
\Delta(m;x) \approx \frac{h}{(2\pi)^d} \sum_{k=-N}^{N} \frac{1}{e^{kh}} \left[\frac{\pi}{z_k+1/\Lambda^2}\right]^{d/2}
\exp\left[-z_km^2 - \frac{x^2}{4(z_k+1/\Lambda^2)}\right],
\end{equation}
\begin{table}
\begin{center}
\begin{tabular}{|c|l|}
\hline 
Exact & 0.654038297612956387655447 \\
\hline
$N=   10$ & 0.6{\sl 86283404027993439164852} \\
   20 & 0.65{\sl 5465233659481763033090} \\
   40 & 0.65403{\sl 7798068595478871392} \\
   80 & 0.65403829761{\sl 3089305988864} \\
  100 & 0.654038297612956{\sl 949544002} \\
\hline
\end{tabular}
\end{center}
\caption{Convergence of Sinc approximation to integral.  (Exact integral
calculated using Maple V, with 40 digits of precision.)  $x^2 = 10$, $m = 1$,
$\Lambda^2 = 10$.}
\label{table:calculation}
\end{table}
The example in Table \ref{table:calculation} demonstrates that we can have as
many digits as are necessary for the calculation, with the associated increase
in compute time.  For practical purposes, 80 terms is appropriate for most
hardware floating-point representations.  Thus we have a form for a two-point
scalar Green function, regulated by the scale $\Lambda^2$ with constant
computational cost regardless of spacetime dimension.  We can take derivatives
explicitly or by construction:
\begin{equation}
\partial^2 \Delta(m;x) = m^2\Delta(m;x) - \bar\delta(x)
\end{equation}
where $\bar\delta(x) = e^{-x^2\Lambda^2/4}$

From this representation, we can directly construct a fermion two-point
function:
\begin{equation}
S(m;x) = (\gamma\cdot\partial - m)\Delta(m;x)
\end{equation}
These representations mean that free scalar and free fermion results
are exact and immediate in any Galerkin evaluation of these trivial
cases.  Furthermore, because of this simplicity, we have the basis for a
complete numerical approach to conventional perturbation theory.

\section{Results with Lehmann representation: $\phi^4$} 
\begin{figure}
\begin{center} \includegraphics[height=2in]{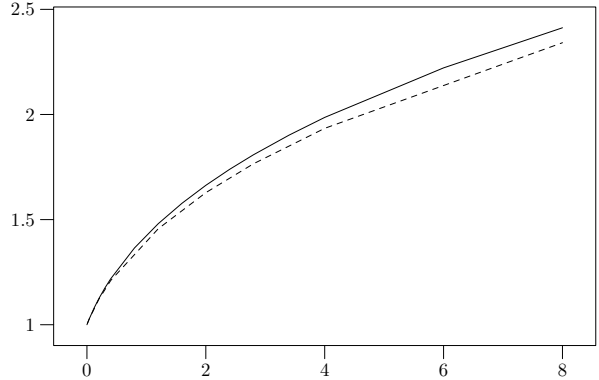}
\caption{One dimensional $\phi^4$ mass gap versus coupling (dashed
line gives exact from Hioe and Montroll, 1975)}
\label{fig:lowest}
\end{center}
\end{figure}

We itemize some results obtained using a regulated single propagator with
parameters set by the Source Galerkin method.   At lowest order, our ans\"atz
for the generating functional is
\begin{equation}
\Z^* = \exp \int \frac{1}{2}j_xG_{xy}j_y.
\end{equation}
Results for this ans\"atz are given in Figure \ref{fig:lowest}.
These results are strikingly accurate and can matched up essentially exactly
with results of Monte Carlo calculations in two and higher dimensions.

\begin{figure}
\begin{center}
\includegraphics{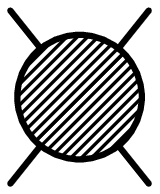}\\
\includegraphics{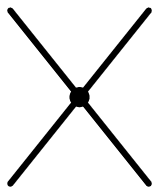}
\includegraphics{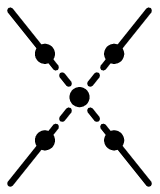}
\includegraphics{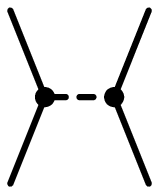}
\includegraphics{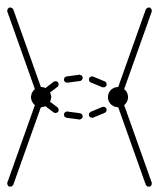}
\end{center}
\caption{Additional connector-based ans\"atzen for the four-point
function, $H$, in $\lambda\phi^4$.  In the bottom row, we have two
contact ans\"atzen on the left, followed by two mediated ans\"atzen.}
\label{figure:phi4-four-point}
\end{figure}

\begin{figure}
\begin{center}
\includegraphics[height=4in]{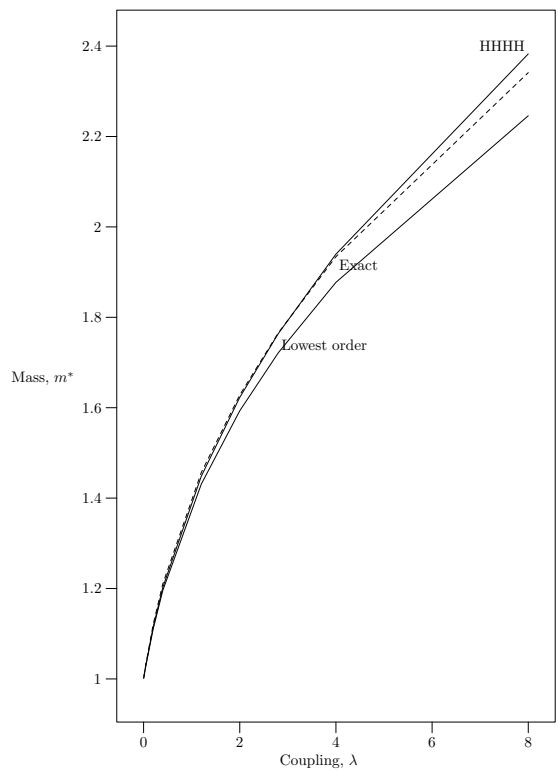}
\end{center}
\caption{Comparison of four-$H$ approximation with lowest order and
exact answer, in one-dimension ($\Lambda^2=70$).} 
\label{p4:fig-hhhh-correction}
\end{figure}

We can enhance these results by including additional 4 source terms in
$\Z^*$. Some simple additional terms that we include with weights and
masses to be calculated using the Source Galerkin technique are the terms of the
forms given in Figure \ref{figure:phi4-four-point}.
The effect of adding a fourth order term is shown in
Figure \ref{p4:fig-hhhh-correction}.

In addition to the illustrations given here, we have examined $(\bar\psi\psi)^2$
in the mean field in two dimensions and have found rapid convergence to the
known results from large-$N$ expansions.  It therefore appears that, at least
for the simple cases studied, we have produced a numerical method which draws on
the structural information already known in general through symmetry and
spectral representations which when combined with Galerkin averaging to set
parameters converges with very simple guesses for the vacuum amplitude to known
correct answers produced through other methods of solution including Monte Carlo
methods.  More complicated gauge problems are under study.

\section{Numerical issues}
In this very brief presentation we have avoided discussion of many of
the difficult numerical issues involved in constructing this 
approach. We note some of these issues here without discussion to have
on record:
\begin{itemize}
\item Both interpolative and spectral problems result in medium- to large-scale
non-linear systems; systems solvable using many variable Newton's method
\item Finite storage is the key constraint for interpolative representations,
which must be constrained to two-point ``connectors'', particularly in high
dimensions
\item Storage is a non-issue for spectral representations (memory use entirely
for caching calculated quantities)
\item Both methods are also time-bound to connector-based
representations for higher point functions
\item Time cost from internal loops; however, algorithm can be made
parallel via partitioning of sums
\item Resolution of elementary pole structures ({\it i.e.\/} differentiating
between $\delta$ and $(p^2+m^2)^{-1}$) may be addressable with
arbitrary precision libraries
\item Arbitrary precision may also be useful for regulated perturbation
calculations.
\end{itemize}
These numerical issues, and related general numerical techniques for source
Galerkin are discussed in greater detail in
\cite{phd:garcia,phd:lawson,phd:hahn}.

\section{Conclusions}
We have discussed a method for numerical calculations for field theories on the
continuum; this method being based on the source Galerkin technique introduced
in \cite{group:initial}.  The direct approach, using Sinc functions for
interpolation, is effective but will ultimately be limited by the finite nature
of current computational resources.  The Lorentz invariant regulated representation
derived from Lehmann representation does not suffer from these limitations, and
is applicable to both perturbation and mean-field-based ans\"atzen.  This
approach has the computational advantages of minimal memory utilization and
parallelizable algorithms and also allows direct representation of fermionic
Green functions.  Finally, a number of useful peripheral calculations can made
using this appoximate representation: one can calculate diagrams in a regulated
perturbation theory, as well as calculating dimensionally regularized loops
numerically.  In general, this technique of evaluating field theories takes
advantage of the symmetries of the Lorentz group; future work includes the
extension of the method to more general internal groups, such as gauge groups or
supersymmetry.

\section*{Acknowledgments} 
This work was supported in part by U. S. Department of Energy grant
DE-FG09-91-ER-40588---Task D.  The authors have been the beneficiaries of many
valuable conversations with S. Garc\'\i a, Z. Guralnik, J. Lawson, K. Platt, and
P. Emirda\u g.  Certain results in this work were previously published in Hahn
\cite{phd:hahn}.  Computational work in support of this research was performed
at the Theoretical Physics Computing Facility at Brown University.


\begin{thebibliography}{1}

\bibitem{phd:garcia}
S.~Garc\'{\i}a.
\newblock {\em A new numerical method for quantum field theory}.
\newblock PhD thesis, Brown University, 1993.

\bibitem{group:initial}
S.~Garc\'{\i}a, G.~S. Guralnik, and J.~W. Lawson.
\newblock A new approach to numerical quantum field theory.
\newblock {\em Physics Letters}, B333:119, 1994.

\bibitem{group:ultralocal}
S.~Garc\'{\i}a, Z.~Guralnik, and G.~S. Guralnik.
\newblock Theta vacua and boundary conditions of the {S}chwinger--{D}yson
  equations.
\newblock {\tt hep-th/9612079}, 1996.

\bibitem{phd:hahn}
S.~C. Hahn.
\newblock {\em Functional methods of weighted residuals and quantum field
  theory}.
\newblock PhD thesis, Brown University, 1998.

\bibitem{phd:lawson}
J.~W. Lawson.
\newblock {\em Numerical method for quantum field theory}.
\newblock PhD thesis, Brown University, 1994.

\bibitem{group:lawson:fermions}
J.~W. Lawson and G.~S. Guralnik.
\newblock New numerical method for fermion field theory.
\newblock {\em Nuclear Physics}, B459:612, 1996.
\newblock {\tt hep-th/9507131}.

\bibitem{group:lawson:bosons}
J.~W. Lawson and G.~S. Guralnik.
\newblock Source {G}alerkin calculations in scalar field theory.
\newblock {\em Nuclear Physics}, B459:589, 1996.
\newblock {\tt hep-th/9507130}.

\bibitem{mono:stenger:sinc-methods}
F.~Stenger.
\newblock {\em Numerical Methods Based on {S}inc and Analytic Functions}.
\newblock Springer Series in Computational Mathematics. Springer--Verlag, 1993.

\end{thebibliography}
\end{document}